\documentstyle[times,graphics,astrobib,amssymb]{mn2e}

\newcommand{\be}{\begin{equation}}
\newcommand{\e}{\end{equation}}
\newcommand{\bear}{\begin{eqnarray}}
\newcommand{\ear}{\end{eqnarray}}

\newcommand{\f}{\frac}
\newcommand{\de}{{\rm d}}

\begin{document}

\title[Reionization sources]
{On the minimum mass of reionization sources}
\author[Choudhury, Ferrara \& Gallerani]
{T. Roy Choudhury$^{1}$\thanks{E-mail: chou@ast.cam.ac.uk},~
A. Ferrara$^{2}$\thanks{E-mail: ferrara@sissa.it}
and
S. Gallerani$^{2,3}$\thanks{E-mail: galleran@sissa.it}
\\
$^{1}$Institute of Astronomy, Madingley Road, Cambridge CB3 0HA, UK\\
$^{2}$SISSA/ISAS, via Beirut 2-4, 34014 Trieste, Italy\\
$^{3}$Institute of Physics, E\"otv\"os University, P\'azm\'any P. 
s. 1/A, 1117 Budapest, Hungary}

\maketitle

\date{\today}

\begin{abstract}
By means of carefully calibrated semi-analytical reionization models, we
estimate the minimum mass of star-forming haloes required to match the current data. 
Models which do not include haloes of total mass $M < 10^9 M_{\odot}$ fail at 
reproducing the Gunn-Peterson and electron scattering optical depths simultaneously, as
they contribute too few (many) photons at high (low, $z\approx 6$) redshift. 
Marginally acceptable solutions require haloes with $ M \approx 5 \times 10^7 M_{\odot}$ 
at $z \approx 10$, corresponding to virial temperatures ($\sim 10^4$K) for which cooling
can be ensured by atomic transitions. However, a much better match to the data is 
obtained if minihaloes ($M \sim 10^6 M_{\odot}$) are included in the analysis. 
We have critically examined the assumptions made in our model and conclude that 
reionization in the large-galaxies-only scenario can remain viable only if metal-free stars 
and/or some other exotic sources at $z > 6$ are included. 
\end{abstract}
\section{Introduction}

Current models of reionization, when compared with 
QSO absorption line measurements and CMB polarization experiments, seem
to indicate that reionization is a complex process extending
over $6 < z < 15$. However, the sources which were primarily responsible
for the process still remain uncertain. Even if one makes
the (not-so-drastic) assumption that reionization is primarily
driven by UV photons from stellar sources, the exact nature of the
stars and the mass range of the hosting galaxies are still open
questions. 

For example, semi-analytical models of
\citeN{cf06b}, which are consistent with a wide variety of
observational data sets,
predict that reionization is mostly driven by haloes of
mass $< 10^9 M_{\odot}$ harboring metal-free
stars at $z \approx 10$ \cite{cf07}. Radiative transfer simulations
of \citeN{imsp07} conclude that the constraints on the electron
scattering optical depth $\tau_{\rm el}$ \cite{sbd++07} 
are satisfied by simply including
haloes above $10^8 M_{\odot}$; no exotic sources or minihaloes are required.
Using a comprehensive model for galaxy formation, \citeN{mlgzd07}
conclude that  the
IGM can be completely reionized at $z \approx 6-7$ by massive stars
within protogalactic spheroids with halo
masses $\sim 10^{10}-10^{11} M_{\odot}$ 
without resorting to any special stellar IMF; such models are also
found to be consistent with the bounds on $\tau_{\rm el}$.
On the other hand, using the observational constraints
on the Ly$\alpha$ optical depth at $z=6$, \citeN{bh07} conclude
that the reionization process is ``photon-starved'' and 
considerable photon contribution at $z > 6$ is required to complete
reionization by $z=6$.
Numerical simulations of \citeN{gnedin07} predict negligibly small
escape of photons from haloes with $M < 10^{11} M_{\odot}$, and
hence it is quite difficult to produce enough photons so as 
to reionize the IGM by $z=6$.
On the observational front, using the 
observed value of the assembled mass at $z \simeq 5$ and 
currently available 
(but highly uncertain) rate of decline in the star formation history over 
$5<z<10$, it can be concluded that a considerable fraction of star-formation
is not yet observed at high redshifts \cite{sbeel07}. This could be either due to significant dust extinction at early times or because of an abundant population of low-luminosity sources just beyond the detection limits of current surveys,
thus implying a reionization scenario by small galaxies.

Given such wide variety of conclusions in the literature, it is important to 
examine in detail the kind of halo masses required to match the available
observational data. In particular, it would be interesting to check whether
models with only large galaxies (say, haloes with masses $> 10^9 M_{\odot}$) 
with standard stellar spectra and IMF are able to match the data, or is there
a desperate need for minihaloes ($M \sim 10^6 M_{\odot}$) and/or metal
free (PopIII) stars or any other exotic source. To address this question,
we use the semi-analytical formalism of \citeN{cf05} and \citeN{cf06b}
(hereafter CF05 and CF06 respectively)
and consider a series of physically-motivated scenarios which differ 
in the minimum mass of star-forming haloes. The main idea of this work
is to confront each of these scenarios with the QSO absorption line
data at $z \approx 6$ and the constraints on $\tau_{\rm el}$ and 
determine if some of the scenarios can be conclusively ruled out.
Throughout the paper, we use the best-fit cosmological parameters from the 3-year WMAP data \cite{sbd++07}, i.e., 
a flat universe with $\Omega_m = 0.24$, $\Omega_{\Lambda} = 0.76$, and $\Omega_b h^2 = 0.022$, and $h=0.73$.  The 
parameters defining the linear dark matter power spectrum are $\sigma_8=0.74$, $n_s=0.95$, $\de n_s/\de \ln k =0$.

\section{Basic features of the model}
The main features of the semi-analytical model used in this work
could be summarized along the following (for a more detailed description
see CF05 and CF06): The model accounts for IGM inhomogeneities by adopting a lognormal distribution with the evolution of volume filling factor of
ionized hydrogen (HII) regions $Q_{\rm HII}(z)$ being calculated 
according to the method outlined in \citeN{mhr00}; 
reionization is said to be complete once all the low-density regions (say, with overdensities $\Delta < \Delta_{\rm crit} \sim 60$) are ionized. 
Hence, the distribution of high density regions determines the 
mean free path of photons
\be
\lambda_{\rm mfp}(z) = \f{\lambda_0}{[1 - F_V(z)]^{2/3}}
\label{eq:lambda_0}
\e
where $F_V$ is the volume fraction of ionized regions and 
$\lambda_0$ is a normalization constant fixed 
by comparing with low redshift observations
of Lyman-limit absorption systems \cite{smih94}.

The number of ionizing photons depends on the assumptions made regarding 
the sources.
In this work, we have assumed two types of reionization sources: 

{\it (i) Stellar sources:} We assume that the photon production rate
from stars within haloes is proportional to the formation
rate of haloes, which in turn is calculated using the 
Press-Schechter formalism. All haloes
above a threshold mass $M_{\rm min}$ are allowed to form stars.
The stellar sources are assumed to have 
metallicities $Z=0.2 Z_{\odot}$ and form with a Salpeter IMF in the mass 
range $1 - 100 M_{\odot}$; the stellar emission spectra are obtained from
the population synthesis models of \citeN{bc03}. 

Under the above assumptions, the characterization of the
stellar sources require only two free parameters as
far as reionization studies are concerned, namely, (i) the efficiency
parameter of stars $\epsilon \equiv \epsilon_* f_{\rm esc}$ where
$\epsilon_*$ is the fraction of baryonic mass within haloes converted
into stars and $f_{\rm esc}$ is
the escape fraction of ionizing photons from the host halo
and (ii) the minimum mass of haloes $M_{\rm min}$ which are able to form stars.
In this work, we assume $\epsilon$ to be independent of redshift and
halo mass, while different physically-motivated models for $M_{\rm min}$ 
are chosen and studied, as will
be discussed in the next section.

Note that the quantity $M_{\rm min}$ introduced above corresponds
to star-forming haloes {\it only within neutral regions}.
Reionization by UV sources is accompanied by 
photo-heating of the gas, which results in a suppression of 
star formation in low-mass 
haloes within ionized regions, a process known as
radiative feedback. Hence, the minimum mass
of star-forming haloes within ionized regions $M_{\rm min}^{\rm RFB}$
could be substantially larger
than $M_{\rm min}$ introduced above.
We compute the value of $M_{\rm min}^{\rm RFB}$ self-consistently from the 
evolution of the gas temperature in the ionized regions and is typically
$\sim 2-3 \times 10^8 M_{\odot}$ at $6 < z < 10$.

Note that we do not include any metal-free (i.e. PopIII) stars, 
which is the main difference of this work compared to our previous works
(CF05, CF06).

{\it (ii) QSOs:}  In this work, we
compute the emissivity of QSOs using likelihood
estimations of the observed luminosity
function at $z < 6$ \cite{meiksin05}. The main uncertainty in the QSO 
contribution comes from the slope of the faint end of the luminosity 
function which is poorly constrained observationally \cite{sw07}. 
In this work, we 
include the contribution of only those QSOs whose luminosities are 
above the break or characteristic luminosity; hence
the QSO contribution should be considered as a lower limit while the
actual emissivity could be a few times higher. Our estimates are similar
to or lower than that of \citeN{meiksin05} and \citeN{bh07}.

\nocite{fsb++06,ksm++06,tkk++06,gffc07,gsfc07}
\begin{figure*}
\rotatebox{270}{\resizebox{0.5\textwidth}{!}{\includegraphics{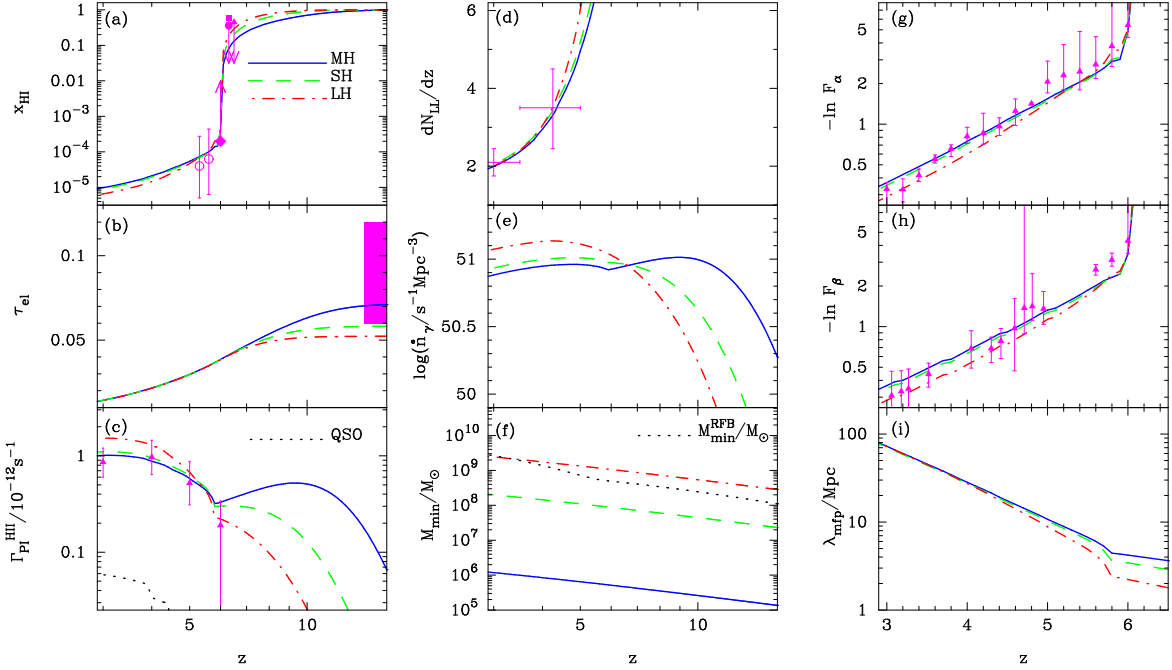}}}
\caption{Comparison of model predictions with observations for
different models described in the text and summarized in Table 1. 
The different panels indicate:
(a): The volume-averaged neutral hydrogen 
fraction $x_{\rm HI}$, with observational
limits from QSO absorption lines (Fan et al. 2006; diamond), 
Ly$\alpha$ emitter luminosity function (Kashikawa et al. 2006; triangle) and
GRB spectrum analysis (Totani et al 2006; square). Also shown
are the constraints using dark gap statistics on QSO spectra 
(Gallerani et al 2007a; open circles) and GRB spectra (Gallerani et al. 2007b;
filled circle).
(b): Electron scattering optical depth, with  observational constraint from
WMAP 3-year data release.
(c): Photoionization
rates for hydrogen, with estimates from numerical simulations (shown
by points with error-bars; Bolton et al. 2005, Bolton \& Haehnelt 2007).
The dotted line shows the lower limit of the QSO contribution.
(d): Evolution of Lyman-limit systems, with observed data points from Storrie-Lombardi et al. (1994). 
(e): Emission rate of ionizing photons per comoving volume. 
(f): The minimum mass of haloes which are allowed to form stars within
neutral regions. The dotted line denotes the corresponding minimum
mass within ionized regions obtained using the radiative
feedback prescription.
(g): Ly$\alpha$ effective optical
depth, with observed data points from Songaila (2004) and Fan et al. (2006). 
(h): Ly$\beta$ effective optical
depth, with observed data points from Songaila (2004)  and Fan et al. (2006).
(i): Evolution of the photon mean free path in physical units.
}
\label{fig:basicplots}
\end{figure*}

The main observational data sets used in this work are those
of the transmitted fluxes $F_{\alpha}$ and $F_{\beta}$
in Ly$\alpha$ and Ly$\beta$ regions respectively, as obtained from
the QSO absorption spectra. We have taken the points tabulated
in \citeN{songaila04} and \citeN{fsb++06}. For calculating
$F_{\alpha}$, we have binned the data points
within redshift intervals of $\Delta z =0.2$ and calculated
the mean. The errors are calculated using 
the extreme values of $F_{\alpha}$
along different lines of sight. 
Hence the errors shown in this paper are typically 
larger than other methods which compute the uncertainties using the
interquartile range \cite{bh07} or standard dispersion. 
For calculating $F_{\beta}$, we note that the data points at $z < 5.5$ are
quite sparse \cite{songaila04} and hence do not require further binning;
we simply use the values and errors tabulated in \citeN{songaila04}. For
points at $z > 5.5$, we follow the method identical to the Ly$\alpha$ case.
The constraints on $\tau_{\rm el}$ are obtained from \citeN{sbd++07}, constraints
on $\Gamma_{\rm PI}^{\rm HII}$ from \citeN{bhvs05} and \citeN{bh07} and the
redshift distribution of Lyman-limit absorption 
systems from \citeN{smih94}.

\vspace{-0.5cm}
\section{Minimum mass of star-forming haloes}
\begin{figure*}
\rotatebox{270}{\resizebox{0.3\textwidth}{!}{\includegraphics{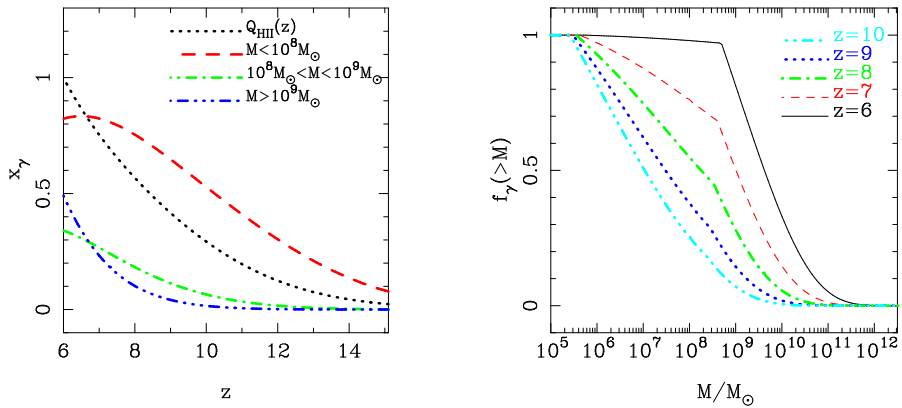}}}
\caption{Photon contribution for the MH model.
{\it Left}: Number of ionizing photons per H-atom (accounting for the number of recombinations) contributed by haloes within a given halo mass range as a function of 
redshift $z$.  The dotted line represents the evolution of the volume filling factor $Q_{\rm HII}$ of ionized regions.
{\it Right}: Cumulative fraction of the ionizing power $f_{\gamma}$ contributed by haloes of mass $>M$. The curves from 
right to left correspond to $z = 6,7,8,9,10$ respectively.
}
\label{fig:minihalo_halofrac}
\end{figure*}

\begin{table}
\caption
{Parameter values for different models used in the paper.}
\begin{center}
\begin{tabular}{|c|c|c|c|c|c|c|}
\hline
Model & $\epsilon=\epsilon_* f_{\rm esc}$\\
\hline
MH & 0.008\\
SH & 0.009\\
LH & 0.013\\
\hline
\end{tabular}
\label{table:modpar}
\end{center}
\end{table}

In this Section, we consider three
physically motivated models which differ in the choice of 
the value of $M_{\rm min}$ and check whether they are able to match
all the data sets. The models are described in the following:

(i) {\it Minihalo (MH)}: The minimum mass of star-forming haloes for this model
is set
by a virial temperature of $T_{\rm vir}=300$K, 
which corresponds to a scenario where molecular cooling is fully efficient. 
Note that $M_{\rm min}$ is redshift-dependent and is typically
$\sim 5 \times 10^5 M_{\odot}$ at $z = 6$. We should mention that 
the star-forming efficiencies of such haloes are debatable \cite{hb06} as H$_2$ could
be easily dissociated by a Lyman-Werner background photons. 
However, it has also been argued that such a background only delay the
star formation in minihaloes and do not necessarily suppress them \cite{on07}.

(ii) {\it Small Halo (SH)}: The minimum mass of star-forming haloes is set
by a virial temperature of $T_{\rm vir}=10^4$K; this is motivated
by the fact that all haloes having $T_{\rm vir} \geq 10^4$K are able
to cool via atomic transitions. This model has 
$M_{\rm min} \sim 10^8 M_{\odot}$ at $z = 6$ and is usually considered
as standard in most semi-analytical works. We must mention again that
for both the SH and MH models, 
the value of $M_{\rm min}$ corresponds to the neutral regions only; 
the minimum mass of star-forming haloes is much larger in ionized regions
because of radiative feedback.

(iii) {\it Large Halo (LH)}: The minimum mass of star-forming haloes is set
by a virial temperature of $T_{\rm vir}=5 \times 10^4$K which corresponds
to $M_{\rm min}(z=6) \sim 10^9 M_{\odot}$. Such value is appropriate to 
a scenario in which reionization is driven by large galaxies which are 
largely unaffected by radiative feedback.

For each model, we find the maximum value of the efficiency $\epsilon$
such that it does not violate the upper bound on 
$F_{\alpha}$ and $F_{\beta}$  at $z = 6$ and then check how it compares with
other observations, in particular whether it can produce 
$\tau_{\rm el} > 0.06$, the 1-$\sigma$ lower limit from
WMAP3 \cite{sbd++07}.
The motivation for normalizing all the models by QSO absorption
line data is that the measurements of $F_{\alpha}$ and $F_{\beta}$ are
less affected by systematics and other uncertainties compared to 
other data sets considered here. In contrast, the
constraints on $\tau_{\rm el}$ obtained from CMB polarization
measurements are still preliminary and the possibility of
major revision in future experiments cannot be ruled out.
Note that, the lower bounds on $F_{\alpha}$ and $F_{\beta}$  at $z = 6$ 
are practically zero and hence the minimum value
of $\epsilon$ cannot be obtained using QSO absorption line
data. However, the
upper bounds should be considered as robust; in fact we have been
quite conservative in this work and used the extreme maximum
value of $F_{\alpha}$ and $F_{\beta}$ allowed by the data.
The values of $\epsilon$ for different 
models after normalizing to the upper limits 
of $F_{\alpha}$ and $F_{\beta}$  at $z = 6$ 
are summarized in Table \ref{table:modpar}.
The results are shown in Figure \ref{fig:basicplots}.

It is clear from the figure that once the models
are normalized to the upper bounds on $F_{\alpha}$ and $F_{\beta}$ at 
$z=6$, the MH and SH models are able to match the evolution
of $F_{\alpha}$ and $F_{\beta}$ up to lower redshifts [Panels (g) and (h)]. 
However, the 
LH model, which does not include low mass ($< 10^9 M_{\odot}$) haloes, 
gives a poor match with the low redshift observations.
Similar conclusions can be drawn from the constraints on 
$\Gamma_{\rm PI}^{\rm HII}$ where the MH and SH models
can fit the data till $z \approx 3$ while the LH model fails
to do so.
More importantly, when compared with the 
observed $\tau_{\rm el}$ [Panel (b)], we note that 
only the MH model can match the data, while
the other two models fall short of the lower 1-$\sigma$ limit.
The LH model predicts $\tau_{\rm el} \approx 0.052$, which can
be considered as a poor match to the data.
The  SH model predicts 
$\tau_{\rm el} = 0.058$, marginally lower than
the 1-$\sigma$ limit; given the uncertainties in the modeling
of the reionization, this could be considered as marginally acceptable.  
Such low value of $\tau_{\rm el}$ is a severe problem for the LH model
because the only way to increase the value of $\tau_{\rm el}$ would be
to increase $\epsilon$ (the only free parameter) which would then
underpredict the Ly$\alpha$ and Ly$\beta$ optical depths at $z=6$.

Hence, models which do not include stars from haloes with masses
$< 10^9 M_{\odot}$ cannot match the GP and 
the electron scattering optical depths simultaneously. In fact, one should
at least include haloes of masses $\sim 5 \times 10^7 M_{\odot}$ to get
a marginal match with the data (the SH model). 
The analysis also brings out the 
importance of including the QSO absorption line measurements explicitly
into any reionization model. For example, a model with only 
$M >  10^9 M_{\odot}$ haloes with a efficiency $\sim 0.06$ would 
reionize the universe around $z \approx 8$ and produce 
$\tau_{\rm el} \approx 0.07$; however it would severely overpredict
the $F_{\alpha}$ and $F_{\beta}$ (and $\Gamma_{\rm PI}^{\rm HII}$ too) 
at $z=6$ and hence
would not be acceptable.

Let us examine which haloes contribute most significantly to reionization; we
shall limit ourselves to the MH model as the other two models
are shown to be unable to match observations.
The number of ionizing photons per H-atom contributed by haloes in the mass range $[M_{\rm min}, M_{\rm max}]$
is given by\footnote{Note that, in our previous work, 
we had defined $x_{\gamma}(z)$ as $\f{n_{\gamma}(z)}{n_H} \f{t_{\rm rec}(z)}{t_H(z)}$, which blows up when recombinations are negligible ($t_{\rm rec} \to \infty$). Under the present definition, $x_{\gamma}(z) \to \f{n_{\gamma}(z)}{n_H}$when $t_{\rm rec} \to \infty$, which is the correct limit.}
\be
x_{\gamma}(z) \equiv \f{n_{\gamma}(z)}{n_H [1 + t_H(z)/t_{\rm rec}(z)]}
\e
where $n_H$ is the comoving number density of hydrogen 
atoms while $n_{\gamma}$
is the time-integrated comoving photon density, calculated using the relation
\be
n_{\gamma}(z) = \int_0^{t(z)} \de t ~ \dot{n}_{\gamma}(M_{\rm min} : M_{\rm max}, t)
\e
where $\dot{n}_{\gamma}(M_{\rm min} : M_{\rm max}, t)$ is the ionizing photon comoving emissivity 
from haloes within $[M_{\rm min}, M_{\rm max}]$.
The term $[1 + t_H(z)/t_{\rm rec}(z)]$ accounts for the number of 
recombinations in the IGM
where $t_H(z)$ is the Hubble time  and
$t_{\rm rec}(z)$ is the recombination time.
By construction, the IGM is reionized when $x_{\gamma} \gtrsim 1$.  
A second quantity of interest is  the fractional instantaneous contribution of 
haloes above a certain mass, 
\be
f_{\gamma}(>M, z) \equiv \f{\dot{n}_{\gamma}(>M, z)}{\dot{n}_{\gamma}(z)}.
\e

The plots of $x_{\gamma}$ and $f_{\gamma}(>M, z)$ for the MH model
is shown in Figure \ref{fig:minihalo_halofrac}. 
It is clear from the figure that 
haloes of mass $< 10^8 M_{\odot}$ dominate the ionizing background
at high redshifts, their contribution decreasing gradually at $z < 8$ because
of radiative feedback. 
However, these haloes are still the dominant contributors of ionizing
photons when integrated till $z=6$ (though the instantaneous 
photon production rate at $z=6$ is dominated by $>10^9 M_{\odot}$
haloes). Hence models which do not include $M < 10^8 M_{\odot}$ haloes
would miss out a large fraction photons at high redshifts (before
radiative feedback is effective) and hence would underpredict
$\tau_{\rm el}$.
For the SH model, we find that
$< 10^8 M_{\odot}$ haloes produce only about 10\% of ionizing photons
when integrated till $z=6$, while about 50\% of photons come
from high mass $>10^9 M_{\odot}$ haloes.

\vspace{-0.5cm}
\section{Discussion}
We have used a semi-analytical reionization model, empirically calibrated on 
a variety of observational data sets, to estimate the minimum mass of ionizing photon 
sources required to match the current data. We find that models which 
do not include haloes with mass $M < 10^9 M_{\odot}$ are not able to reproduce
the GP and electron scattering optical depths {\it simultaneously}. Such models (i)
contribute too few photons at high redshift,  and (ii) {\it produce too many photons 
too late}. To get a marginally acceptable match with the data, one requires
haloes with masses as small as $5 \times 10^7 M_{\odot}$ at $z \approx 10$, 
which would correspond to a virial temperature of $\sim 10^4$K. In such cases, 
though the bulk of photons ($\sim 90\%$) is produced by $M > 10^8 M_{\odot}$ haloes, 
the low mass haloes are important to contribute to $\tau_{\rm el}$
at high redshifts without violating the QSO absorption line constraints
at $z=6$.

A much better match to the data is obtained if minihaloes ($M \sim 10^6 M_{\odot}$)
are included in the analysis. 
These haloes produce enough photons at high redshifts to give a 
high $\tau_{\rm el}$. They are also easily destroyed once radiative
feedback becomes substantial and hence give no contribution to the 
photoionization rate at $z \approx 6$, thus agreeing with the 
$F_{\alpha}$ and $F_{\beta}$ upper bounds.
In case the minihaloes are not allowed to form stars because
of some photodissociating Lyman-Werner background, 
it becomes almost impossible to construct reionization models
with standard stellar sources that are not in tension with data. 
Given this, it is crucial to critically examine the assumptions and idealizations
made in our formalism which could allow reionization scenarios with
only large galaxies to be consistent with the data, which is done in the
following:

(i) $z$-dependence of the photon production efficiency: in this work, 
we have assumed the
efficiency parameter $\epsilon$, the stellar IMF and the stellar spectrum
to be independent of $z$. In case the value of $\epsilon$ was higher
at high redshifts, it could, in principle, produce high $\tau_{\rm el}$ at
high redshifts without violating the GP constraints at $z=6$. Such behavior
of $\epsilon$ would mean that either stars were forming more efficiently
at early times and/or the escape fraction of photons was higher. A similar
effect could also be achieved if the stellar IMF was top-heavy
at high-$z$ or the spectra of the stars were harder. In short, one would
require  a very efficient production of photons per baryons at high-$z$. An obvious
candidate for achieving such effects would be the inclusion of metal-free 
(PopIII) stars with or without a top-heavy IMF. Such models with 
PopIII stars are found to be an excellent match to a wide variety of data sets
in the SH scenario (CF06), while they can possibly be tuned to match the data
in the LH case too.

(ii) Mass-dependence of $\epsilon$: similarly neglected here, is the possibility that
the efficiency parameter depends on the halo mass.
Note that, in order to make the LH scenario work, one would require $\epsilon$ to be higher
for smaller mass haloes so that the photon contribution
increases at $z > 6$. However, the mass-dependence, if any, is found
to be opposite, e.g., \citeN{khw++03} found that $\epsilon$ increases
with halo mass for $M < 3 \times 10^{12} M_{\odot}$
in the local Universe.  Given this, it is unlikely that a 
mass-dependent $\epsilon$ would improve the performance of large-galaxy-only models. 

(iii) Radiative feedback: One of the main uncertainties in theoretical
models of reionization is the implementation of radiative (i.e. photoionization) feedback. However,
note that this effect mostly affects haloes of masses $<10^9 M_{\odot}$ [Panel (f) of Figure \ref{fig:basicplots}] 
and hence a different feedback prescription would have {\it no} effect on the LH model at all. For the SH model, 
a less severe feedback mechanism, which allows the $10^8 M_{\odot} < M < 10^9 M_{\odot}$ haloes to survive 
longer than what is used here \cite{gnedin00}, could produce enough photons at high-$z$ to get a better match to the data. 
On the other hand, if the feedback is more severe on the $\sim 10^8 M_{\odot}$ haloes
(e.g., because of the photoionization rate boost arising from the clustering of galaxies, and not taken 
into account here), the SH model would be ruled out.

(iv) QSOs: The contribution of QSOs considered here should be thought
of as a lower limit; the actual contribution could be much higher. However,
this does not affect our conclusions because a higher contribution
from QSOs at $z \sim 6$ would imply a lower value of $\epsilon$, which would
then produce a much lower $\tau_{\rm el}$.

(v) IGM inhomogeneities: The density distribution of the IGM has been
assumed to be lognormal, which is found to be a good match
to the QSO transmitted flux distribution \cite{gcf06}. However, it has been
argued that the density distribution obtained from simulations 
has a different form \cite{mhr00}. 
The density distribution can affect the results in three ways, 
namely: (a) The evolution of $Q_{\rm HII}(z)$ could be altered
if the density distribution is different; however note that
there is not much freedom observationally in the qualitative
behavior of $Q_{\rm HII}(z)$ as QSO absorption line data
requires reionization to be completed around $z \gtrsim 6$. 
(b) The evolution of $\lambda_{\rm mfp}$ could be different thus
modifying the photoionization rate $\Gamma_{\rm PI}^{\rm HII}$
which is discussed in the next point.
(c) For a given $\Gamma_{\rm PI}^{\rm HII}$, a different
density distribution would give
a different the value of $F_{\alpha}$ (and $F_{\beta}$). However, note that
the analysis presented in the paper could also be done using the
constraints on $\Gamma_{\rm PI}^{\rm HII}$ without any reference to
$F_{\alpha}$ or $F_{\beta}$, and the results would still
be qualitatively similar.

(vi) Photon mean free path: A related problem is that regarding the
value of $\lambda_{\rm mfp}$ at $z=6$. There are no observational
constraints on $\lambda_{\rm mfp}$ at $z > 4$, and the theoretical
estimates would depend on the density distribution of the IGM. In case
$\lambda_{\rm mfp}$ is found to be lower than that obtained in our models
($\lambda_{\rm mfp}(z=6) = 3.73$ and 2.37 proper Mpc for SH and LH respectively), 
it would give a lower $\Gamma_{\rm PI}^{\rm HII}$ for the same value
of $\epsilon$, and hence could allow the SH and LH models to match with
observations. However, the typical values of $\lambda_{\rm mfp}$ found 
using the density distribution of \citeN{mhr00} 
are $\sim 5$ physical Mpc \cite{bh07,wbh07}, which
would clearly rule out the SH and LH models. A trivial extrapolation
of the observed $\lambda_{\rm mfp}$ at lower redshifts to high-$z$
would too give similar values. 

(vii) Revised observational constraints: A good chance of 
the large galaxies scenario to survive (without including PopIII stars
or other sources) would be to revise the 
constraints on $\tau_{\rm el}$. We have already seen that the value of
$\tau_{\rm el}$ was lower in the WMAP3 data release than in the WMAP1
because of systematics. In case the value of $\tau_{\rm el}$ is found
to be $\sim 0.05$, it would be enough to allow the LH scenario. On the
other hand, in case the upper bounds on $F_{\alpha}$ and $F_{\beta}$ are
tightened with increase in QSO sample size, it could rule out the 
LH (and possibly SH) scenario with a higher degree of confidence. 
For example, we have
been conservative in estimating the errors and allowed 
a $F_{\alpha}$ as high as 0.0125 at $z=6$. One should compare this
with the constraints $F_{\alpha} < 0.004$ used by \citeN{bh07}; such
severe constraints would clearly disfavor the LH and SH scenarios.
Another possibility is that the constraints on 
the cosmological parameters are revised, e.g., the value of 
$\sigma_8$ is found to be higher than what is used. A rigorous
exploration of the cosmological parameter space is beyond the
scope of this work. However, a model with higher value of $\sigma_8=0.9$
\cite{vhl06}, 
when normalized to Ly$\alpha$ and Ly$\beta$ flux at $z=6$, 
gives $\tau_{\rm el} \approx 0.059$; this value is still well below
the corresponding 1-$\sigma$ bound on $\tau_{\rm el} \approx 0.1\pm 0.03$.

In spite all the model uncertainties, it
seems certain that reionization with large galaxies scenario 
($M > 10^9 M_{\odot}$) can be conclusively
ruled out with the present data; such scenarios can only be allowed if 
metal-free stars or other exotic sources at high redshifts
are included. 
The scenario where only those haloes which can cool
via atomic transitions contribute is marginally acceptable.
In any case, there seems to be a requirement
for a large number of sources at $z \approx 10$, which are most
likely faint (i.e., low-mass) haloes.
Observationally, it 
is important to put constraints on star formation within
these faint galaxies at high redshift which, however, seems to be a 
challenging task. Nonetheless one should be optimistic as 
most of such issues would be addressed with future experiments like JWST.
On the theoretical front, it is important to realize that reionization
models could be incomplete unless they are compared with both the $\tau_{\rm el}$ and GP constraints simultaneously.

\vspace{-0.5cm}
\section*{Acknowledgement}
SG acknowledges the support by the Hungarian National Office for
Research and Technology (NKTH), through the Pol\'anyi Program.

\vspace{-0.5cm}
\bibliography{mnrasmnemonic,astropap-mod,reionization}
\bibliographystyle{mnras}

\end{document}